\documentclass[preprintnumbers,prd,twocolumn,nofootinbib,
superscriptaddress,nobibnotes,showpacs]{revtex4}
\usepackage{amsfonts}
\usepackage{mathrsfs}
\usepackage{epsfig}
\usepackage{graphicx}%
\usepackage{dcolumn}
\usepackage{amsmath}

\begin{document}

\title{The median recoil direction as a WIMP directional detection signal}


\author{Anne M. Green}
%\email[]{anne.green@nottingham.ac.uk}
%\thanks{}
\affiliation{School of Physics and Astronomy, University of 
  Nottingham, University Park, Nottingham, NG7 2RD, UK}
\author{Ben Morgan}
%\email[]{ben.morgan@warwick.ac.uk}
%\thanks{}
\affiliation{Department of Physics, University of Warwick, Coventry, 
CV4 7AL, UK}


%\date{\today}

\begin{abstract}
  Direct detection experiments have reached the sensitivity to detect
  dark matter WIMPs. Demonstrating that a putative signal is due to
  WIMPs, and not backgrounds, is a major challenge however.  The
  direction dependence of the WIMP scattering rate provides a
  potential WIMP `smoking gun'.  If the WIMP distribution is
  predominantly smooth, the Galactic recoil distribution is peaked in
  the direction opposite to the direction of Solar motion.  Previous
  studies have found that, for an ideal detector, of order 10 WIMP
  events would be sufficient to reject isotropy, and rule out an
  isotropic background. We examine how the median recoil direction
  could be used to confirm the WIMP origin of an anisotropic recoil
  signal.  Specifically we determine the number of events required 
  to confirm the direction of solar motion as the median inverse recoil direction at 95\% confidence.
   We find that for zero background 31 events are
  required, a factor of $\sim 2$ more than are required to simply
  reject isotropy. We also investigate the effect of a non-zero
  isotropic background.  As the background rate is increased the
  number of events required increases, initially fairly gradually and
  then more rapidly, once the signal becomes subdominant.  We also
  discuss the effect of features in the speed distribution at large
  speeds, as found in recent high resolution simulations, on the
  median recoil direction.

\end{abstract}

%\keywords{}

\maketitle

%\section{}
\section{Introduction}

Weakly Interacting Massive Particles (WIMPs), and in particular the
lightest neutralino in supersymmetric models, are a well motivated
dark matter candidate~\cite{jkg,bhs,dmbook}. WIMPs can be detected
directly, in the lab, via the elastic scattering of WIMPs on detector
nuclei~\cite{DD}. Experiments now have the sensitivity required to
probe the theoretically favoured regions of parameter
space~\cite{cdms,xenon10,zeplin} and the CDMS experiment has recently
observed two events in its WIMP signal region~\cite{cdms}.

Neutrons, from cosmic-ray induced muons or natural radioactivity, can
produce nuclear recoils which (on an event by event basis) are
indistinguishable from WIMP induced recoils. Furthermore perfect
rejection of other backgrounds is impossible, for instance `surface
events' (electron recoils close to the detector surface) in the case
of CDMS. As highlighted by the CDMS events, as well as the
long-standing DAMA annual modulation signal~\cite{dama}, demonstrating
the WIMP origin of a putative signal is absolutely crucial. The
direction dependence of the WIMP scattering rate (due to the Earth's
motion with respect to the Galactic rest frame)~\cite{dirndep}
provides a potential WIMP `smoking gun'. Assuming that the WIMP
distribution is predominantly smooth, the peak WIMP flux comes from
the direction of solar motion (towards the constellation Cygnus) and
the recoil rate is then peaked in the opposite direction.  The recoil
rate, in the Galactic rest frame, is highly anisotropic; the rate in
the forward direction is roughly an order of magnitude larger than
that in the backward direction.  A detector capable of measuring the
nuclear recoil vectors (including the sense +${\bf p}$ versus -${\bf
  p}$) in 3-dimensions, with good angular resolution, could reject
isotropy of the recoils with only of order 10
events~\cite{copi:krauss,pap1}.  Most, but not all, backgrounds would
produce an isotropic Galactic recoil distribution (due to the
complicated motion of the Earth with respect to the Galactic rest
frame).  An anisotropic Galactic recoil distribution would therefore
provide strong, but not conclusive, evidence for a Galactic origin of
the recoils.

Confirmation of the WIMP origin could be obtained by verifying that
the properties of the anisotropy match the expectations for WIMP
induced recoils. Assuming the WIMP distribution is dominated by a
smooth component, the median inverse recoil direction should be
compatible with the direction of solar motion.

To summarise, a WIMP search strategy with a directional detector could be divided into a sequence of phases:\\
1. Search phase (detection of non-zero recoil signal)\\
 2. Detection of anisotropy \\
 3. Study of properties of anisotropy \\
which require successively larger numbers of events (and hence larger exposures).

The initial simple search phase aims to detect an anomalous recoil
signal above that expected from backgrounds. To claim an anomalous
signal inconsistent with zero at 95\% confidence requires 4-5
events. The second step, as discussed above, would be to check whether
the Galactic recoil directions are anisotropic and would require of
order 10 events (assuming zero background).  In this paper we focus on
the third phase, examining how measuring the median recoil direction
could be used to provide confirmation of the WIMP origin of an
anisotropic recoil signal.  In section ~\ref{model} we describe the
input to our Monte Carlo simulations.  In section~\ref{res} we present
our results, before concluding with discussion in Sec.~\ref{dis}.

\section{Modelling}
\label{model}

We use the same statistical techniques and methods for calculating the
directional nuclear recoil spectrum as in Refs.~\cite{pap1,pap2,pap3}.
We briefly summarise these procedures here, for further details see
these references and Ref.~\cite{bm:thesis}. 

\subsection{Detector}

Many of the directional detectors currently under
development~\cite{sm,cygnus} are low pressure gas time projection
chambers (TPCs), e.g.  DMTPC~\cite{dmtpc},
DRIFT~\cite{drift,sean:drift} and NEWAGE~\cite{newage}. We therefore
simulate a fairly generic TPC based detector.  We assume that the the
recoil directions, including their senses, are reconstructed perfectly
in 3d. These are optimistic assumptions, therefore our results provide
a lower limit on the number of events/exposure required by a real TPC
based detector. For concreteness we use a $S$ target with an energy
threshold of 20 keV.  Finite angular resolution does not significantly
affect the number of events required to detect
anisotropy~\cite{pap1,copi2d}, provided it is not worse than of order
$10^{\circ}$.  Axial and/or 2-d read-out would, however, significantly
degrade the detector capability~\cite{pap1,pap2,copi2d,pap3}.

\subsection{WIMP properties and distribution}

The detailed angular dependence of the recoil rate depends on the
exact form of the WIMP velocity
distribution~\cite{copi:krauss,pap1,alenazi}. However, if the WIMP
velocity distribution is dominated by a smooth component the main
features of the recoil distribution (rear-front asymmetry, median
direction opposite to the direction of solar motion) are robust (see
e.g. Ref.~\cite{pap1})).  For concreteness we use the standard halo
model halo, an isotropic sphere with local density $\rho=0.3 \, {\rm
  GeV} \, {\rm cm}^{-3}$ and a Maxwellian/gaussian velocity
distribution with three dimensional velocity dispersion $\sigma_{\rm v}=270 \, {\rm km \,
  s}^{-1}$, and fix the WIMP mass at $m_{\chi}=100 \, {\rm GeV}$.

Numerical simulations find velocity distributions with stochastic
features at large speeds~\cite{vogelsberger,kuhlen}.  Kuhlen
et. al.~\cite{kuhlen} find that for high speed WIMPs ($v> 500 {\rm km
  \, s}^{-1}$) the direction in which the WIMP flux is largest can
deviate by more than $ 10^{\circ}$ from the direction of solar
motion. The effect on the median recoil direction will be
substantially smaller than this however. Firstly the set-up we are
considering ($S$ target with energy threshold of 20 keV and
$m_{\chi}=100 \, {\rm GeV}$) is sensitive to WIMPs with $v>225 \, {\rm
  km s}^{-1}$. Therefore a high speed feature will only contribute a
small fraction of the WIMP flux. This will be true in general unless
the WIMP mass is small and/or the threshold energy is large. Secondly,
see e.g.  fig. 3 of Ref.~\cite{pap1}, due to the elastic scattering
the recoil rate is less anisotropic than the WIMP flux. The deviation
of the inverse median direction from the direction of solar motion
will therefore be substantially smaller than the deviation of the peak
WIMP flux.  It will also depend on the (a priori unknown) velocity and
density of the feature. For instance, for a stream with velocity, in
Galactic coordinates, ${\bf v}_{\rm str}= (-65, 135, -249)\, {\rm km
  \, s}^{-1}$ the difference between the inverse median direction and
the solar direction only exceeds $5^{\circ}$ if the fraction of the
local density in the stream exceeds $10\%$~\cite{pap1}.

Since the effect of a feature in the speed distribution on the median
recoil direction is expected to be small we do not investigate it in
this study. If, with a large number of events, a statistically
significant deviation of the inverse median direction were found, its
origin could be investigated by studying the energy dependence of the
deviation.  We defer an investigation of this to future work.

Finally, direct detection experiments probe the ultra-local dark
matter distribution on scales many orders of magnitude smaller than
those resolved by simulations. It is not (and may never be) possible
to directly calculate, or otherwise measure, the dark matter
distribution on the relevant scales. In this case if WIMPs are
detected, then the directional recoil rate could be used to
reconstruct the ultra-local dark matter distribution~\cite{radon}.

\section{Results}
\label{res}

The recoil rate is peaked in the direction opposite to the direction
of solar motion.  To allow ease of comparison with the direction of
solar motion we use the inverse recoil directions (i.e. the directions
from which the recoils originate) in our analyses.

We first examine how the accuracy with which the median Galactic recoil
direction is determined depends on the number of WIMP events.  The median
direction is defined as the direction ${\bf x}_{\rm med} $ which
minimises the sum of the arclengths between ${\bf x}_{\rm med}$ and
the individual inverse recoil directions ${\bf x}_{\rm
  i}$~\cite{fisher:lewis:embleton}. It is found by minimising
\begin{equation}
\label{M}
{\cal M} = \sum_{i=1}^{N} {\rm cos}^{-1} ({\bf x}_{\rm med}.{\bf x}_{\rm i}) \,,
\end{equation}
where $N$ is the number of events.

For a given number of WIMP events, $N_{\rm wimp}$, we simulate $10^4$
experiments and in each determine the direction, ${\bf x}_{\rm med}$,
by minimizing Eq.~(\ref{M}) and hence calculate $\Delta$,
\begin{equation}
\Delta= {\rm cos}^{-1}({\bf x}_{\rm med}.{\bf x}_{\odot}) \,,
\end{equation}
the angle between the median direction and the direction of solar
motion, ${\bf x}_{\odot}$.  In Fig. 1 we plot the 50\% and 95\%
percentiles of the distribution of $\Delta$ as a function of $N_{\rm
  wimp}$.  We also investigate the effects of non-zero (isotropic)
backgrounds. We parametrize the background rate in terms of the
fraction of events which are signal, $\lambda=S/(S+B)$ where $S$ and
$B$ are the signal and background rates respectively
(c.f. ~Ref.~\cite{billard}). For comparison we also plot the $5\%$ and
$50\%$ percentiles for an purely isotropic distribution.

\begin{figure}
\includegraphics[width=8.5cm]{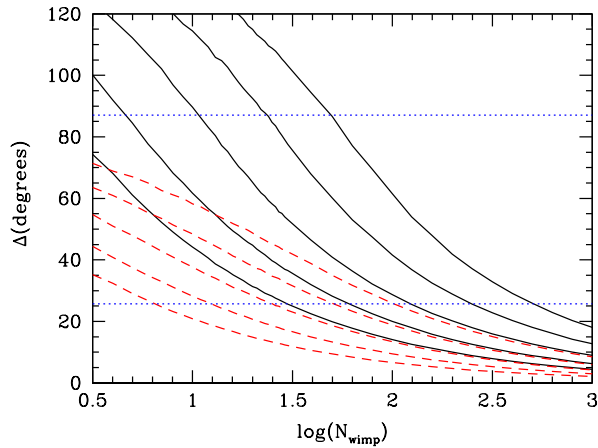}
\caption{The distribution of the angle between the median inverse
  recoil direction and the direction of solar motion, $\Delta$, as a
  function of the number of WIMP events, $N_{\rm wimp}$, for varying
  signal fractions (from top to bottom) $\lambda=0.0625, 0.125, 0.25,
  0.5$ and $1$. The solid and dashed lines are the $95\%$ and $50\%$
  percentiles.  The dotted lines are (from top to bottom) the $50\%$
  and $5\%$ percentiles for an isotropic recoil distribution.}
\label{deltaN}
\end{figure}

\begin{figure}
\includegraphics[width=8.5cm]{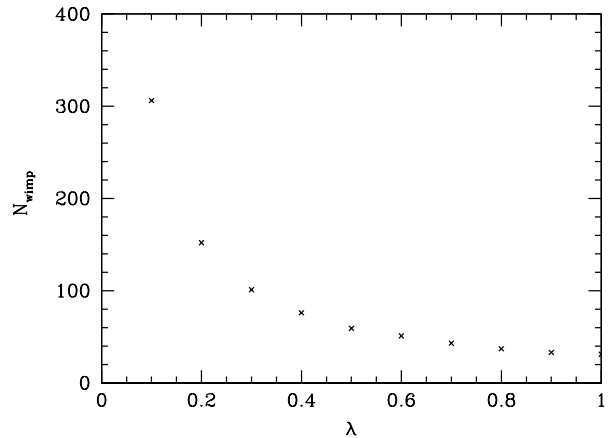}
\caption{The dependence of the number of WIMP events, $N_{\rm wimp}$, required to reject the
  median direction being random at 95$\%$ confidence in $95\%$ of
  experiments on the signal fraction, $\lambda$.}
\label{Nlambda}
\end{figure}

We now determine the number of events required  to confirm the direction of solar motion as the median inverse recoil direction at 95\% confidence. We do this using the same methodology as in
Ref.~\cite{pap1}. Briefly, we use the distribution of $\Delta$ for
WIMP induced recoils and for the null hypothesis of isotropic recoils,
to calculate the rejection and acceptance factors, $R$ and $A$. The
rejection factor gives the confidence level with which the null
hypothesis can be rejected given a particular value of the test
statistic, while the acceptance factor is the probability of measuring
a larger value of the test statistic if the alternative hypothesis is
true, We then find the number of WIMP events required to give
$A=R=0.95$ i.e. to reject the median direction being random at 95$\%$
confidence in $95\%$ of experiments.  This is shown in
Fig.~\ref{Nlambda}, as a function of the signal fraction
$\lambda$. For zero background (i.e. $\lambda=1$) 31 events are
required, a factor of $\sim 2$ more than are required to simply reject
isotropy. The number of events required increases as $\lambda$ is
decreased, initially fairly gradually and then, once the signal
becomes subdominant ($\lambda \lesssim 0.5$), more rapidly.

Billard et al.~\cite{billard} have recently investigated using a
sky-map based likelihood analysis to probe the correlation of the
directional recoil rate with the direction of solar motion (and hence
confirm the WIMP origin of a signal). Their results are qualitatively
consistent with ours.  For instance they find that, for a signal
fraction $\lambda=0.5$, the peak signal direction can be confirmed to
be within $20^{\circ}$ of the direction of solar motion with as few as
25 WIMP events. The median direction method is however faster and more
robust.  It particular it has the advantage of being model independent
(i.e. one does not need to assume an exact form of the WIMP velocity
distribution).

If desired, further confirmation of the WIMP origin of an anisotropic
recoil signal could be obtained by studying the variation of the
median recoil direction in the lab frame~\cite{drift}. Over the course
of a sidereal day the peak in the recoil distribution traces out a
small circle on the sky. For a detector located in the Northern
hemisphere during a sidereal day the peak recoil direction in the lab
rotates (roughly) from down to South and back. A periodogram analysis
could be used to verify that median direction varies over a sidereal,
rather than solar, day (c.f. Ref.~\cite{amperiod} for the annual
modulation).

\section{Discussion}
\label{dis}

In this paper we have investigated how the median recoil direction in
directional detection experiments can be used as a WIMP
signal. Assuming a smooth WIMP distribution, the peak WIMP flux is
from the direction of solar motion and the median recoil direction is
in the opposite direction.  An ideal detector could reject isotropy of
recoils with only of order 10
events~\cite{copi:krauss,pap1}. Confirmation of the WIMP origin of an
anisotropic recoil signal could be obtained by studying the details of
the anisotropy and in particularly confirming that the median inverse
Galactic recoil direction coincides with the direction of Solar
motion. We find that, with an ideal detector and zero-background, to
confirm the direction of solar motion as the median inverse recoil direction at 95\% confidence 
requires 31 events (see also
Ref.~\cite{billard}). Non-zero isotropic background would increase
this number, significantly if the signal is sub-dominant.
  
For concreteness (and in the absence of a definitive alternative) we
have used the standard halo model halo, which has an isotropic
Maxwellian speed distribution.  While the detailed angular dependence
of the recoil rate depends on the exact form of the WIMP velocity
distribution, if the WIMP velocity distribution is dominated by a
smooth component the median inverse recoil distribution will be close
to the direction of solar motion (see e.g. Ref.~\cite{pap1}).  Recent
high resolution simulations have found stochastic, features in the
speed distribution at large speeds~\cite{vogelsberger,kuhlen}.  The
direction in which the flux of high speed WIMPs is largest can deviate
by more than $ 10^{\circ}$ from the direction of solar
motion~\cite{kuhlen}. However the deviation of the median inverse
direction will be small compared with that expected from an isotropic
recoil direction. We therefore conclude that features in the speed
distribution at high speed are unlikely to affect the utility of the
median recoil direction as a WIMP signal.  If, with a large number of
events, a statistically significant deviation of the inverse median
direction were found, its origin could then be investigated by
studying the energy dependence of the deviation.

\begin{acknowledgments}
  AMG and BM are supported by STFC. 
 \end{acknowledgments}


\end{document}